\documentclass[onecolumn,12pt]{IEEEtran}

\usepackage{amssymb,mathtools}
\usepackage{graphicx, times, amsmath, amsfonts, comment,amsthm}
\usepackage[noend]{algorithmic}
\usepackage{algorithm, array}
\usepackage{epstopdf}
\usepackage{tikz}
\begin{document}

\title{Deep Learning for Radio Resource Allocation in Multi-Cell Networks}

\author{K. I. Ahmed,  H. Tabassum, and E. Hossain\thanks{K. I. Ahmed and E. Hossain are with the Department of Electrical and Computer Engineering at the University of Manitoba, Canada. H. Tabassum is with the Department of Electrical Engineering and Computer Science at York University, Canada.  
}}

\maketitle

\begin{abstract}
Increased complexity and heterogeneity of emerging 5G and beyond 5G (B5G) wireless networks will require a paradigm shift from traditional resource allocation mechanisms. Deep learning (DL) is a powerful tool where a  multi-layer neural network can be trained  to model a resource management algorithm  using network data.Therefore, resource allocation decisions can be obtained without intensive online computations which would be required otherwise for the solution of resource allocation problems.  In this context, this article focuses on the application of DL to obtain solutions for the radio resource allocation problems in multi-cell networks. Starting with a brief overview of a deep neural network (DNN) as a DL model, relevant DNN architectures and the data training procedure, we provide an overview of existing state-of-the-art  applying DL in the context of radio resource allocation. A qualitative comparison is provided in terms of their objectives, inputs/outputs, learning  and data training methods. Then, we present a supervised DL model to solve the  sub-band and power allocation problem in a multi-cell network.  Using the data generated by a genetic algorithm, we first train the model and then test the accuracy of the proposed model in predicting the resource allocation solutions. Simulation results show that the trained DL model is able to provide the desired optimal solution $86.3\%$ of time.

\end{abstract}

\begin{IEEEkeywords}
5G and B5G cellular, radio resource allocation, deep learning, deep neural networks, auto-encoder, genetic algorithm, prediction accuracy.
\end{IEEEkeywords}

\section*{Introduction}

Optimal radio resource allocation is one of the fundamental challenges for design and operation of cellular wireless networks.  Typically, the resource allocation problems are often formulated and solved using tools from optimization theory. These problems need to be solved for specific network scenarios taking into account the rapidly varying wireless channels and quality-of-service (QoS) requirements of the users. When the optimization problems are non-convex (which is often the case in a real-world scenario), the optimal solutions are obtained by applying exhaustive search methods, genetic algorithms, combinatorial, and branch and bound techniques which incur significantly high time and computational complexities. Therefore, these methods are not appealing for large-scale heterogeneous cellular networks with ultra-dense base station (BS) deployments, massive connections, and different resources (e.g. transmission time, frequency channels, antennas, transmit power) and diverse QoS requirements for different classes of users. Sub-optimal solutions obtained based on techniques such as Lagrangian relaxations, iterative distributed optimization, heuristic algorithms, and cooperative game theory are also often very computation intensive and/or may not be feasible for  large cellular networks due to high signaling overhead. Also, these sub-optimal solutions can be far from optimal solutions and their convergence properties and the optimality gap could be unknown. 

Machine learning (ML) tools can be used  to obtain practical solutions for radio resource allocation problems in a large cellular network given the past optimal or near-optimal resource allocation decisions. Compared to iterative distributed optimization and heuristic sub-optimal techniques, ML-based resource allocation algorithms can be implemented online. With ML, the required information is learned directly from the data samples instead of complicated mathematical models.  ML automatically extracts complex features from data samples which is difficult for humans.  Nevertheless, the traditional ML problems become difficult when the environments are heterogeneous, data collected from various sources have different formats,  data demonstrate complex correlations, and have high dimensions. In such scenarios, significant human intervention is required to make accurate inferences and decisions based on the data, which is referred to as {\em feature engineering}. Deep learning (DL), which is a sub-class of ML, hierarchically extracts high-level features and correlations from input data without  human efforts through multiple layers of nonlinear processing units.  DL models can handle a large amount of data (as expected in 5G/B5G networks due to massive wireless connections), which improves the prediction accuracy as well, and also unlabeled data can  be used for learning important patterns in an unsupervised manner. DL reduces computational and time complexity since a trained model can achieve multiple objectives without the need of retraining and Graphics Processing
Unit (GPU)-based parallel computing enables DL to make inferences within milliseconds~\cite{goodfellow2016deep,hinton2006reducing}. 




With the DL approach for radio resource management in a large cellular network, a multi-layer neural network model can be built to which the relevant training samples can be provided as inputs, and then the resource allocation solution can be obtained as outputs without intensive computations (as each layer performs simple operations like matrix-vector multiplications~\cite{slhina1}).  Furthermore, with the advent of software defined networking (SDN) and cloud storage technologies, it is possible to deal with the storage, analysis, and computation of {\em big data}  as well as training the DL models on  SDN controllers. 



In this article, we first review the basic concepts of a deep neural network (DNN) architecture as a DL model and the training procedure for the DL model.  Then, we review the existing literature DL-based resource allocation problems.  To this end,
we present a supervised DL model to compute optimal sub-band and power allocation solutions for a multi-cell network (e.g. a cloud-RAN) with an objective to maximizing the total network throughput. This is a well-known non-convex combinatorial optimization problem. As such, to generate near-optimal training data for the DL model, we utilize a genetic algorithm (GA). Simulation results show that the proposed model can obtain the desired solutions for sub-band and power allocation with 85\% accuracy.  

\begin{figure*}[t]
\begin{center}
\includegraphics[width=5in]{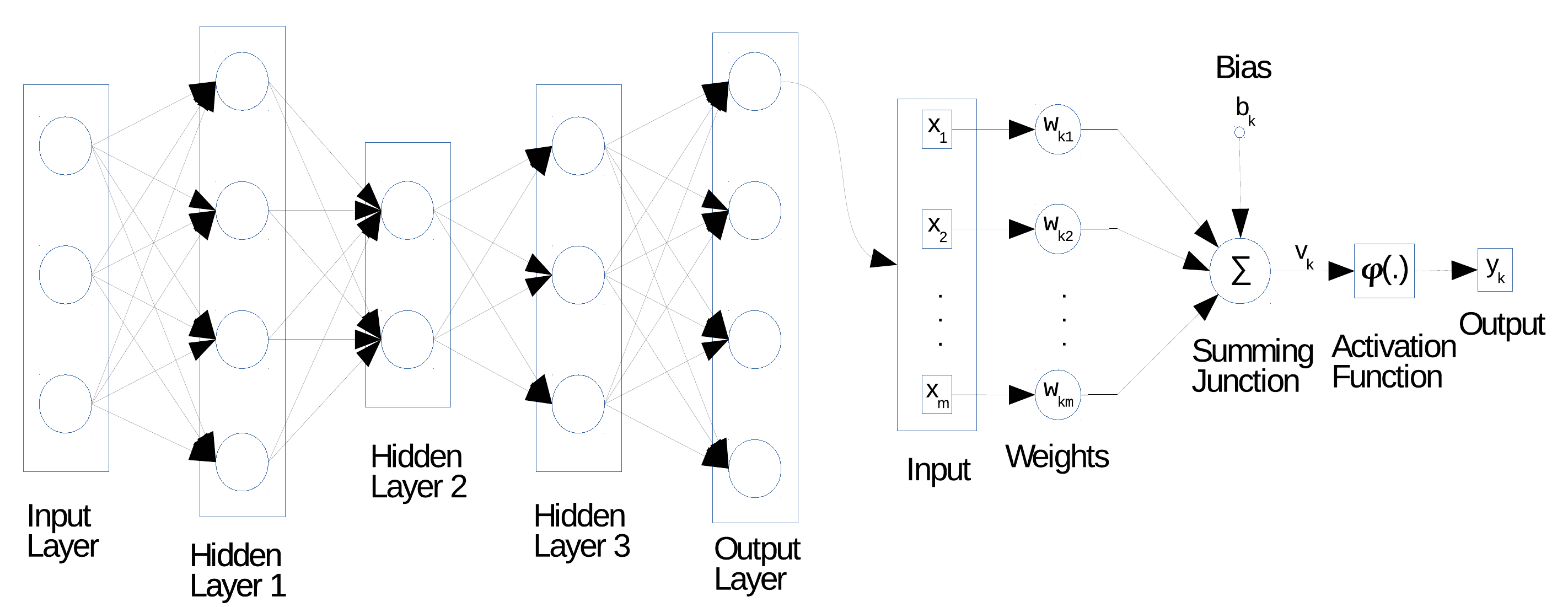}
\end{center}
\caption{A deep neural network with three hidden layers.}
        \label{fig_deep}
\end{figure*}

\section*{Fundamentals of Deep Learning}
DL is a specific methodology to train and build neural networks. A basic DL model is a multi-layered feed-forward neural network that enables feature extraction and transformation of input data. We briefly review the basic composition of deep neural networks (DNN), relevant DNN architectures and data training procedure with feed-forward and back propagation DNN.

\subsection*{Deep Neural Network (DNN)}

A DNN consists of an input layer, multiple hidden layers, and the output layer (Fig.~\ref{fig_deep}). Each layer has multiple units commonly known as neurons. Each neuron performs a non-linear operation on the inputs. A weighted summation of the inputs is first computed, and then before sending the weighted summation to an activation function (e.g. a Sigmoid function $\varphi(x) =  \frac{1}{1+e^{-x}}$), a  bias value is added. The activation function creates a non-linear relationship between the input and the output~\cite{goodfellow2016deep}. 
The non-linear representation of a neuron is shown in Fig.~\ref{fig_deep}. Every neuron has a vector of weights and a bias value. Therefore, for a layer, there will be a  weight vector and a bias vector. If $\mathbf{x}$ is the input of  hidden layer $i$, then the output of the hidden layer $i$ is
$ \mathbf{h}^{(i)}=\varphi^{(i)}(\mathbf{w}^{(i)T}\mathbf{x} + \mathbf{b}^{(i)})$,
where $\varphi^{(i)}$ is the activation function, $\mathbf{w}^{(i)}$ is the weight vector, and $\mathbf{b}^{(i)}$ is the bias vector of hidden layer $i$.

\subsection*{Deep Learning Architectures}
DL relies on four primary neural network architectures: unsupervised pre-trained networks (UPNs), convolutional neural networks (CNNs), recurrent neural networks (RNN), and recursive neural networks. The DL architectures can be used along with three learning models: supervised, unsupervised, and reinforcement learning. In supervised DL, a supervisor helps the model to learn the features from data, i.e. the model already knows the output of the algorithm before it starts learning it. Therefore, in supervised learning, the dataset contains the target for each input. Supervised learning is used in classification and regression problem. In unsupervised learning, the model tries, and there is no supervisor to provide correct answers. Therefore, in unsupervised learning, the dataset does not have any target. And the primary goal is to correctly infer the outputs for a given set of unlabeled input data. In reinforcement learning (a popular example of which is {\em Q-learning}), there is a software agent who learns by interacting with the environment. The agent senses its current state and the state of the environment and then choose an action. For every action it takes, there is a consequence. The agent either receives a reward for a good move or a penalty for a bad move. The primary job of the agent is to maximize the cumulative reward through a series of actions. When a neural network is used as an agent,  it is referred to as deep reinforcement learning (DRL).

In this article, we focus on {\em Auto-encoder} (AE) which falls under the category of UPNs. An AE  is an unsupervised learning model \cite{baldi2012autoencoders} which reconstructs the inputs at the outputs. By doing so, the AE learns the feature space of the data set. It normally consists of an input layer with one or more hidden layers, and an output layer. The number of neurons in the input layer and the output layer should always be the same. An AE consists of (i) an encoder which transforms the input to a hidden code and (ii) a decoder which reconstructs the input from the hidden code. 
There are various types of AEs, namely, denoising AE, stacked AE, sparse AE, and variational AE. For instance, a sparse AE can learn sparse features, i.e. a small number of hidden neurons will respond to a specific feature of the data set. By adding a sparse penalty into the traditional AE model, i.e. a constraint to force the average activation of hidden neurons to be zero (close to zero), the performance of the AE can be significantly improved~\cite{sun2016sparse}. Another variant is the stacked AE which consists of multiple layers of sparse AEs. Usually, greedy layer-wise training is employed for stacked AE.

\subsection*{Data Training Procedure}
Let us assume that we have to learn a target function  $ \mathbf{y}= f^{*}(\mathbf{x})$. Here, $\mathbf{x}$ is the input vector and $\mathbf{y}$ is the output vector. The DL model is therefore $ \mathbf{y}= f(\mathbf{x}; \mathbf{\theta})$, where $\mathbf{\theta}$ denotes  the unknown parameters, i.e.  weights and biases. Our goal is to learn $\mathbf{\theta}$ precisely so that our model can be closer to the original one. Different algorithms (e.g.  Stochastic Gradient Descent (SGD), Momentum, Nesterov Momentum, AdaGrad, RMSProp, and Adam) can be used to learn $\mathbf{\theta}$ \cite{goodfellow2016deep}. 

To learn $\mathbf{\theta}$,  a training dataset composed of inputs and outputs is typically used to train the model. In the training phase, we first initialize the weights and biases randomly. Then, we feed those inputs to the input layer of the system. The output of the input layer is used as an input for the hidden layer. In this way, the data propagates through the hidden layers and finally reaches the output layer. The propagation of information from the input layer to the output layer is known as the {\em forward pass}. A cost function (e.g. mean squared error [MSE]) is used to measure the quality of the model by calculating the error between the predicted and the original value. 
The resulting error signal is then propagated backward through the hidden layers and updates the weights and biases of each layer which is known as the {\em backward pass}.  This training process continues until the error rate reaches a threshold value. When training data completes a forward and backward pass, we call it a complete training cycle or epoch. The training process can also be terminated if it reaches the maximum number of epochs. Such a neural network is referred to as  a {\em feed-forward and back-propagation neural network} and is used in our proposed DL model.


\section*{Deep Learning-Based Wireless Resource Allocation: State-of-the-Art}


DL techniques have recently been used in a variety of wireless resource management problems (e.g. channel and power allocation, throughput maximization, spectrum sharing). We categorize the major works into  two groups: deep reinforcement learning (DRL) and supervised DL. A summary of these works is provided in Table~\ref{table:summary} for a qualitative comparison among these works in terms of the inputs and outputs, data training and generation method used, radio resources considered, objectives of resource management, and their  limitations.

\subsection*{Deep Reinforcement Learning (DRL)}

In \cite{xu2017deep}, the authors use a DRL approach for allocating resources to remote radio heads in a cloud-RAN in order to minimize the total power consumption while ensuring the demand of each user. The model defines the state space, action space, and reward function for the DRL agent. A DNN is used to approximate the action-value function and a convex optimization problem is solved with a reduced action space. The proposed resource allocation scheme is not centrally optimized. 
In~\cite{rlhina1}, a decentralized sub-band and power allocation problem for a vehicle-to-vehicle (V2V) communication system is solved based on DRL. Each V2V link operates as an agent making its own allocation decisions optimally (to minimize interference under latency constraints). In \cite{challita2017proactive}, the authors model the throughput maximization problem of a small cell network with unlicensed spectrum as a non-cooperative game and propose a DL-based solution. More specifically,  a DRL algorithm is developed based on a long short-time memory (LSTM) network which is a particular type of deep RNN for channel selection, carrier aggregation, and fractional spectrum access. The primary objective is to maximize throughput in each small cell while maintaining fairness with co-existing networks. In \cite{8352517}, the authors develop a deep Q-learning based spectrum sharing approach for primary and secondary users in a non-cooperative fashion. The primary users use a fixed power control strategy while the secondary users learn autonomously to adjust the transmission power to share the common spectrum. 




\subsection*{Supervised Learning}

In \cite{lee2018deep}, the authors propose a distributed power control method based on a data-driven convolutional neural network (CNN) which exploits the spatial features in channel gain to estimate the transmit power. However, there is no guarantee that the proposed method can give the centrally optimized solution. In \cite{zapponeonline}, the authors develop an energy-efficient power control scheme based on a trained DNN which takes the communication channels as input and predicts the transmit power. 
In \cite{slhina3},  the authors formulate a weighted throughput maximization problem and solved it using a recurrent DNN architecture. The model is trained by applying the concept of universal function approximation of DNN and Lagrangian duality. That is, 
the non-convex  problem is formulated as a finite-dimensional unconstrained optimization problem in the dual domain with bounded sub-optimality. 
Primal-dual descent methods as well their zeroth-ordered equivalents are used to perform data generation. In \cite{slhina2}, throughput maximization of a device-to-device (D2D)  network is considered with maximum power constraint of a D2D transmitter and maximum interference received by cellular BSs. The input data is generated through simulations. The accuracy/quality of the learning solutions is, however, not described clearly.
Also, in \cite{slhina1}, a throughput maximization problem has been considered with power allocation variables in interference-limited wireless networks using supervised DL by generating data through approximate optimal solutions. 


\begin{table*}[]
\centering
\caption{Summary of existing works on resource allocation using deep learning.}
\label{table:summary}
\begin{tabular}{|l|l|l|l|l|l|l|}
\cline{1-7}
 \textbf{Publication}& \begin{tabular}[t]{@{}l@{}}\textbf{Learning} \\ \textbf{Method} \end{tabular}  & \textbf{Model}  &\textbf{Objective} & \textbf{Resources}& \textbf{Input} & \textbf{Approach} \\ \cline{1-7}
 
 \begin{tabular}[t]{@{}l@{}}Xu {\em et al.} \\\cite{xu2017deep} \end{tabular}   & Reinforcement & Q-learning  & \begin{tabular}[t]{@{}l@{}} Minimize the total power consumption\\ with users' QoS constraints. \end{tabular} &  \begin{tabular}[t]{@{}l@{}}Power \\ \end{tabular} & \begin{tabular}[t]{@{}l@{}} Demand of\\ the users \end{tabular}& \begin{tabular}[t]{@{}l@{}}Centralized\\ \end{tabular}\\ \cline{1-7}
 
 \begin{tabular}[t]{@{}l@{}}Li {\em et al.} \\\cite{8352517} \end{tabular}   & Reinforcement & Q-learning  & \begin{tabular}[t]{@{}l@{}} Spectrum sharing with power control\\  at secondary user \end{tabular} &  \begin{tabular}[t]{@{}l@{}}Power \\ \end{tabular} & \begin{tabular}[t]{@{}l@{}}SINR \end{tabular}& \begin{tabular}[t]{@{}l@{}}Distributed \end{tabular}\\ \cline{1-7}
 
 \begin{tabular}[t]{@{}l@{}}Ye {\em et al.} \\\cite{rlhina1} \end{tabular}   & Reinforcement & Q-learning  & \begin{tabular}[t]{@{}l@{}} Interference minimization  with latency\\ constraints \end{tabular} &  \begin{tabular}[t]{@{}l@{}}Channel, \\Power \\ \end{tabular}&\begin{tabular}[t]{@{}l@{}}Instantaneous \\ channel \end{tabular} & \begin{tabular}[t]{@{}l@{}}Distributed \end{tabular}\\ \cline{1-7}
 
\begin{tabular}[t]{@{}l@{}}Challita {\em et al.} \\\cite{challita2017proactive} \end{tabular}   & Reinforcement & \begin{tabular}[t]{@{}l@{}} RNN \end{tabular}  & \begin{tabular}[t]{@{}l@{}} Maximize throughput with  fairness\\ constraints \end{tabular} &  \begin{tabular}[t]{@{}l@{}}Channel, \\ Time \end{tabular} & \begin{tabular}[t]{@{}l@{}}Unlicensed \\channel \end{tabular} & Distributed \\ \cline{1-7}

\begin{tabular}[t]{@{}l@{}}Lee {\em et al.} \\\cite{lee2018deep} \end{tabular}   & Supervised & \begin{tabular}[t]{@{}l@{}} CNN \end{tabular}  & \begin{tabular}[t]{@{}l@{}} Maximize throughput through  power \\control  \end{tabular} &  \begin{tabular}[t]{@{}l@{}}Power \\ \end{tabular} & \begin{tabular}[t]{@{}l@{}}Normalized\\ channel \end{tabular}& \begin{tabular}[t]{@{}l@{}}Both centralized\\ and distributed \end{tabular}\\ \cline{1-7}

\begin{tabular}[t]{@{}l@{}}Zappone {\em et al.} \\\cite{zapponeonline} \end{tabular}   & Supervised & \begin{tabular}[t]{@{}l@{}}DNN \end{tabular}  & \begin{tabular}[t]{@{}l@{}} Maximize the energy efficiency \\through power control \end{tabular} &  \begin{tabular}[t]{@{}l@{}}Power \\ \end{tabular} & \begin{tabular}[t]{@{}l@{}} Channel \end{tabular}& \begin{tabular}[t]{@{}l@{}}Centralized \end{tabular}\\ \cline{1-7}

\begin{tabular}[t]{@{}l@{}}Eisen {\em et al.} \\\cite{slhina3} \end{tabular}   & Supervised & \begin{tabular}[t]{@{}l@{}}RNN \end{tabular}  & \begin{tabular}[t]{@{}l@{}} Weighted throughput\\ maximization \end{tabular} &  \begin{tabular}[t]{@{}l@{}} Power \\ \end{tabular}&\begin{tabular}[t]{@{}l@{}}Channel gain \end{tabular} &\begin{tabular}[t]{@{}l@{}} Centralized \end{tabular}\\ \cline{1-7}

\begin{tabular}[t]{@{}l@{}}Kim {\em et al.} \\\cite{slhina2} \end{tabular}   & Supervised & \begin{tabular}[t]{@{}l@{}}DNN \end{tabular}  & \begin{tabular}[t]{@{}l@{}} Throughput maximization\\with power and interference constraints \end{tabular} &  \begin{tabular}[t]{@{}l@{}}Power \\ \end{tabular}&\begin{tabular}[t]{@{}l@{}} Unclear \\ \end{tabular} &\begin{tabular}[t]{@{}l@{}}Distributed \\ \end{tabular}\\ \cline{1-7}

\begin{tabular}[t]{@{}l@{}}Sun {\em et al.} \\\cite{slhina1} \end{tabular}   & Supervised &  \begin{tabular}[t]{@{}l@{}}DNN \end{tabular} & \begin{tabular}[t]{@{}l@{}} Throughput maximization \end{tabular} &  \begin{tabular}[t]{@{}l@{}}Power \\ \end{tabular}& \begin{tabular}[t]{@{}l@{}} Channel\\ coefficients \end{tabular} &\begin{tabular}[t]{@{}l@{}}Centralized \end{tabular}\\ \cline{1-7}

\end{tabular}
\end{table*}


\section*{sub-band and Power Allocation in Multi-cell Networks: A Supervised Deep Learning Approach}

The existing reinforcement learning-based methods do not leverage the data obtained for optimal/near-optimal solutions and attempt to find the optimal solutions through trial and error.  Data-driven DL solutions are rather scarce. The existing few studies generally focus on single variables only (e.g. power) and consider heuristic algorithms or simulations to generate the data and train the model.  Therefore, the learning solutions are not efficient solutions.  In the following, we develop a DL-based resource allocation model with an objective to maximizing the total network throughput by performing joint resource allocation (i.e. both power and channel). We use a supervised learning approach to train the DNN model and obtain the training data by solving the non-convex optimization problem through a genetic algorithm.  We also use the stacked auto-encoder approach to pre-train the model. We observe that a pre-trained model converges more quickly compared to an untrained model and performs better.

\subsection*{System Model}
We consider a downlink  cellular network of $K$ base stations (BSs). Each BS $k \in \{1, \cdots, K\}$ has  $F$ frequency sub-bands. The bandwidth of each sub-band is $B$ MHz. The power allocated by cell $k$ in frequency sub-band $f$ is ${P}_{k,f}$ which is discrete. The total power of a cell $k$ is limited by a maximum value $P_{k}^{\max }$ such that $ \sum _{f\in F} P_{k,f} \leq P_{k}^{\max },\quad \forall k\in K$.
Let $\mathcal {U}_{k}$ denote the set of users who are associated with cell $k$, and $\mathcal{U}$ is the set of all users in the network. The vector $\mathcal{A}_{k,f}$ denotes the sub-band allocation on sub-band $f$ in cell $k$, where $\mathcal{A}_{k,f}$ is an integer value.  

The utility for each cell $k$ is defined as follows: 
\begin{equation}\label{U}
U =\sum _{k\in \{1, \cdots, K\}}\sum _{u\in \mathcal {U} _{k}} \sum _{f=1}^{F} 
\left [{ \mathbb{I}(\mathcal{A}_{k,f}=u)  B \log \left ({1 + \alpha {\textrm {SINR}}_{u,k,f}}\right )}\right ]
\end{equation}
where  $\alpha $ is a constant for a given target Bit Error Rate (BER) which is defined as $\alpha = -1.5 / \log (5\textrm{BER}) $. We assume $\textrm{BER}$ to be $10^{-6}$. The signal-to-interference-plus-noise ratio (SINR) of user $u$ when served by cell $k$ which transmits over frequency sub-band $f$ is expressed as
$ {\textrm {SINR}}_{u,k,f} = \frac {P_{k,f} G_{u,k,f}} {\eta _{u} +\sum _{l \neq k} P_{l,f} G_{u,l,f}}$.
where $\eta _{u}$ represents the receiver noise and $G_{u,k,f}$ denotes the link gain from cell $k$ to user $u$ over frequency sub-band $f$  defined as 
$
    G_{u,k,f}=10^{-(PL_{u}+X_{\alpha})/10}.|H_{u,k,f}|^2,
$
where $H_{u,k,f}$ is the Rayleigh fading gain of user $u$ from cell $k$ over frequency sub-band $f$, $X_{\alpha}$ is the log-normal shadowing, and $PL_{u}$ is the path-loss of user $u$.

All users periodically send their channel quality as a channel quality indicator (CQI) to their nearest BS,   where
$\mathcal{C}_{u,k}$ is the CQI vector of a user $u$ of cell $k$ over all frequency sub-bands. That is, $\mathcal{C}_{u,k}$ is a  vector of discrete values   $\mathcal{C}_{u,k} \mathrel {\mathrel {\mathop :}\mkern -1.2mu=}(\mathcal{C}_{u,k,1}, \mathcal{C}_{u,k,2},\ldots, \mathcal{C}_{u,k,F})$. In addition, users are also classified into cell-center users and cell-edge users depending on the users' locations. A location indicator $\mathcal {V}_{u,k}$ is used to indicate whether a user $u$ of cell $k$ is cell-center user or cell-edge user as follows:
\begin{equation} 
\label{eq_location}
\mathcal {V}_{u,k} = 
\begin{cases}
1 & \mbox {if}~~{{R}_{u,k}> R/2} \\
0 & \mbox {otherwise} \end{cases} 
\end{equation}
where ${R}_{u,k}$ is the distance of user $u$ in cell $k$ from the BS and $R$ is the cell radius. For each user class, sub-band and power allocations will be learned and predicted separately.

\subsection*{Supervised Deep Learning Approach}
\begin{figure*}[t]    
\centering 
        \includegraphics[width=5in]{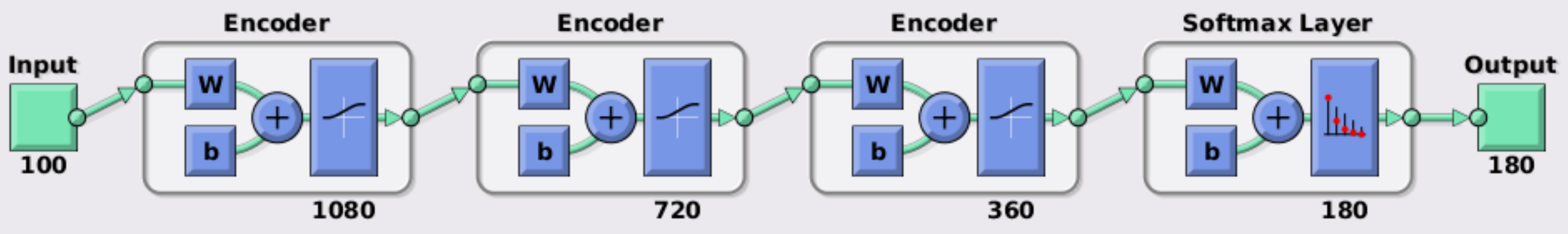}
        \caption{DNN model for power and sub-band allocation.}
        \label{fig_deep1}
\end{figure*}

Now we present a deep learning approach that can predict optimal sub-band and power allocation solutions for multi-cell networks with a certain accuracy. Specifically, this deep learning model takes $\mathcal{C}_{u,k}$ vector along with $\mathcal {V}_{u,k}$ of all users in a network as input and predict the powers and sub-band allocations as output. That is, for $K$ cells, $U$ users and $F$ sub-bands,  the size of the input data is $(K \times U\times (F+1))$. We convert the output data from decimal base to $n$-bits binary base and use their complement in the output also. Therefore, for $K$ cells, the size of  output data is $(2\times 2\times n\times K \times F)$. The approach proceeds in the following phases:

\begin{itemize}

\item\textit{Data generation:} It requires the optimal solution of the sub-band and power allocation problem which is a non-convex combinatorial optimization problem.
One way is to check all possible combinations of power and channel allocation for all the BSs which is referred to as {\em exhaustive search}. For example, with $15$ cells, $5$ sub-bands, and $5$ discrete power levels, then there will be $5^5$ or $ 3125$ possible combinations available for the power setting of one BS. Therefore, we will need to check $3125^{15}$ combinations, which is practically infeasible.  As such, to generate data to train and test our DNN model, we resort to a GA, which is a heuristic searching algorithm inspired by the theory of natural evolution~\cite{sivanandam2008genetic}.  A GA has the following five phases:
\begin{itemize}
    \item \textit{Initial population:} A set of randomly initialized individuals called the initial population is generated. Gene is a group of variables which is normally used to characterize the individual. A stack of genes is called Chromosome.     
    \item \textit{Fitness function:} The fitness function calculates a fitness score for every individual. 
    \item \textit{Selection:} Two sets of individuals (parents) are selected based on their fitness scores. An individual with high fitness score has a high chance of getting selected. 
    \item \textit{Crossover:} A randomly generated crossover point is used for each pair of parents to exchange their genes among themselves to create new offspring. Then the new offspring is added to the population.   
    \item \textit{Mutation:} Some genes of the new offspring, can be mutated with a low mutation probability. The genes are mutated to maintain diversity within the population. 
\end{itemize}
GA continuously generates new offsprings until the population becomes converged, i.e. a new offspring is not significantly improved from the previous generation.  Then we have a set of solutions for our problem. 

In our model, the specific steps are as follows:

\noindent\textbf{Step 1:} 
Allocate a random power vector for each cell. After that, we calculate the CQI value of each sub-band for each user in the network. 
We also estimate the location indicator for every user using Eq.~\ref{eq_location}. Note that the CQI values and the location indicator for every user are the inputs of our DNN model.

\noindent\textbf{Step 2:} Find the power vector of every BS which maximizes the total network utility (Eq.~\ref{U}).  We use GA to solve this problem as explained in the following.

\textbf{Step 2.1:} Generate random power vectors for every cell. This is known as the {\em initial population} of GA. 

\textbf{Step 2.2:} Using the power vectors, we compute the total network utility (Eq.~\ref{U}), which refers to a {\em fitness function} in a GA. Therefore, for every individual, we have a fitness score. 

\textbf{Step 2.3:} We choose two sets of individuals of high fitness score from the initial population. 

\textbf{Step 2.4:} We then choose a random crossover point for each pair and exchange the genes among them. Here, genes are the different power levels of the power vector. 

\textbf{Step 2.5:} With a very low mutation probability, we randomly choose one or more genes of the new offspring and change their values.

\textbf{Step 2.6:} We check the fitness score of the new offspring. If the fitness score improved from their parents, we would put them on the population. We keep repeating \textbf{Steps 2.2-2.6} until there is no significant improvement in the fitness scores. 

\noindent\textbf{Step 3:} Once we find the optimal power vector, we need to calculate the sub-band allocation. A sub-band will be allocated to that user who will maximize the throughput. The power vector and allocation vector are the outputs.

\noindent\textbf{Step 4:} We need to repeat \textbf{Steps 1-4} until we have a certain amount of data to train our DNN model. 

Once data generation is complete, we train our model with the data.  Finally, we use the pre-trained DNN model to predict sub-band and power allocations for every BS in the network. This prediction will be performed online.  

\item \textit{Training phase:} 
The training dataset consists of input data and labeled target data or the output data. We use a stacked AE to initialize the weights and biases for the neurons in the hidden layers. In this way, we pre-train our model, and then we add a Softmax layer to map the input with the target data. A Softmax layer assigns probabilities to each class in such a way that the total probability must be $1.0$.  Finally, we stack the encoder part of the AE with the Softmax layer and fine-tune the model with our training dataset. The complete DNN is shown in Fig.~\ref{fig_deep1}
\item \textit{Testing phase}: After training the model, we test the model on a new dataset and calculate the accuracy of the model.  In a practical setting, all users in the network periodically send their CQI values to their serving BSs, which extract the CQI value of each sub-band and add a location indicator, i.e. cell-center user or cell-edge user. Therefore, for every user, there will be a vector of CQI and location indicator. Each BS then sends the processed information of all users to a central entity (e.g. SDN controller), which runs the DNN model. The DNN model will generate the allocation vector and power vector for all the BSs. Once the prediction is made, the controller will send back the allocation and power vectors to their designated BSs. 


\end{itemize}


\section*{Simulations and Results}

We consider a network of $K$ = 5 BSs with a coverage radius $R=500$ m, maximum transmit power = $40$ W, directional antenna, number of users per cell = $5$, bandwidth of a sub-band $B$ = 2.88 MHz, white noise power density = $-174$ dBm/Hz, number of sub-bands = $3$, number of power levels = $3$, and power per frequency sub-band = $\{ 6.4, 12.8, 19.2\}$ W. 

\subsection*{Input Data Generation}
First, we generate data for the training of DNN model using GA approach. To confirm the accuracy of the solutions obtained from GA, we compare GA with the exhaustive search. For this,  we first simulate a network of $4$ macro-cells where each cell has $5$ users. Then using \textbf{Step 1}, we calculate the CQI values and the location indicator of each user in the network. For the exhaustive search, instead of \textbf{Steps 2 and 3}, we perform the exhaustive search to maximize the total network utility, i.e. the optimal powers.  We repeat \textbf{Steps 1-4} to generate 1000 samples for comparison between the exhaustive search and GA. The comparisons are shown in Tab.~\ref{table:2}.

\begin{table}[]
\centering
\caption{Performance comparison between exhaustive search and genetic algorithm}
\label{table:2}
\begin{tabular}{|l|l| l| l l}
\cline{1-3}
 \textbf{Parameter}& \textbf{Exhaustive Search}  & \textbf{Genetic Algorithm} &  &  \\ \cline{1-3}
Avg. Execution time & 1460.00 sec.  & 118.76 sec. &  &  \\ \cline{1-3}
Max. Execution time  & 2247.50 sec. & 158.65 sec.  & &  \\ \cline{1-3}
Min. Execution time  & 1509.00 sec. & 95.13 sec. &  &  \\ \cline{1-3}
Accuracy  & 100\%  & 85.25\% &  &  \\ \cline{1-3}
\end{tabular}
\end{table}

An exhaustive search takes much longer time compared to GA. This is because, in an exhaustive search, we need to check the whole solution space whereas GA takes a much shorter time and generate optimal solutions with a probability of 0.85 (i.e. prediction accuracy is 85\%).  For a large scale system, it is not feasible to generate data using the exhaustive search. Therefore, we use GA to create training and testing data for our DNN model. We use a network of $5$ BSs where the cells are partially overlapped, and five users are located randomly within each macro-cell. We apply $3$ frequency sub-bands and $3$ power levels for each macro-cell. By using \textbf{Steps 1-4},  we produce around $17000$ samples for our DNN model. Then we use $80\%$ of this data-set for training and the remaining $20\%$ for testing purposes. 

\subsection*{Training the DNN}
After data generation, we train our DNN model with our training data-set. The training data-set has two parts: input and output.  The input data size is $5\times 5\times (3+1)=100$ and for the output data, we convert the data from decimal base to $3$-bits binary base. We also use their complements in the output, i.e. if the binary representation $3$ is ${(011)}_b$, then its complement is ${(100)}_b$. Then, the output data size becomes $2\times 2 \times 3\times 5\times 3= 180$. We use a stacked AE model for pre-training as well as to build our DNN model. We first train an AE with our input data. The first AE has the following training parameters: Transfer function= ``sigmoid", sparsity proportion=$0.15$, sparsity regularization= $4$, L2 weight regularization= $0.004$, maximum number of epochs  = $1000$. Then we generate hidden codes of the first AE by using the input of the training data as an input of the AE. Then these hidden codes are used to train the second auto-encoder. Using this approach, we pre-train several AEs, and at the end, we add a Softmax layer. To pre-train the Softmax layer, we use the hidden codes of the last AE as the input and output part of the training data-set as target or output. 

\subsection*{Results and Discussions}
\subsubsection*{Impact of the number of hidden layers on prediction accuracy} 
\begin{figure}[!htb]
\minipage{0.4\textwidth}
  \includegraphics[width=\linewidth]{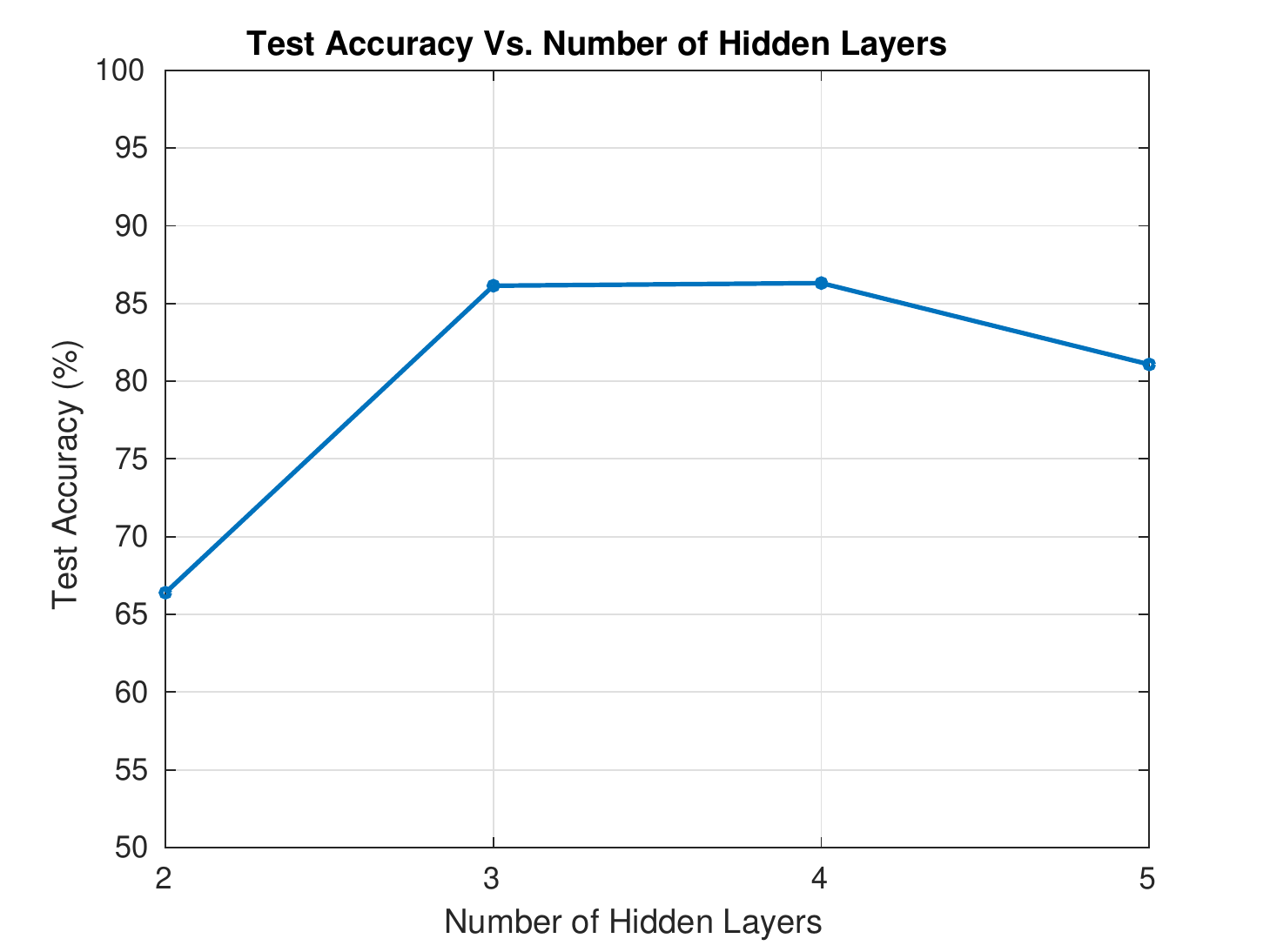}
  \caption{Test accuracy vs. number of hidden layers.}\label{fig:4}
\endminipage\hfill
\minipage{0.4\textwidth}
  \includegraphics[width=\linewidth]{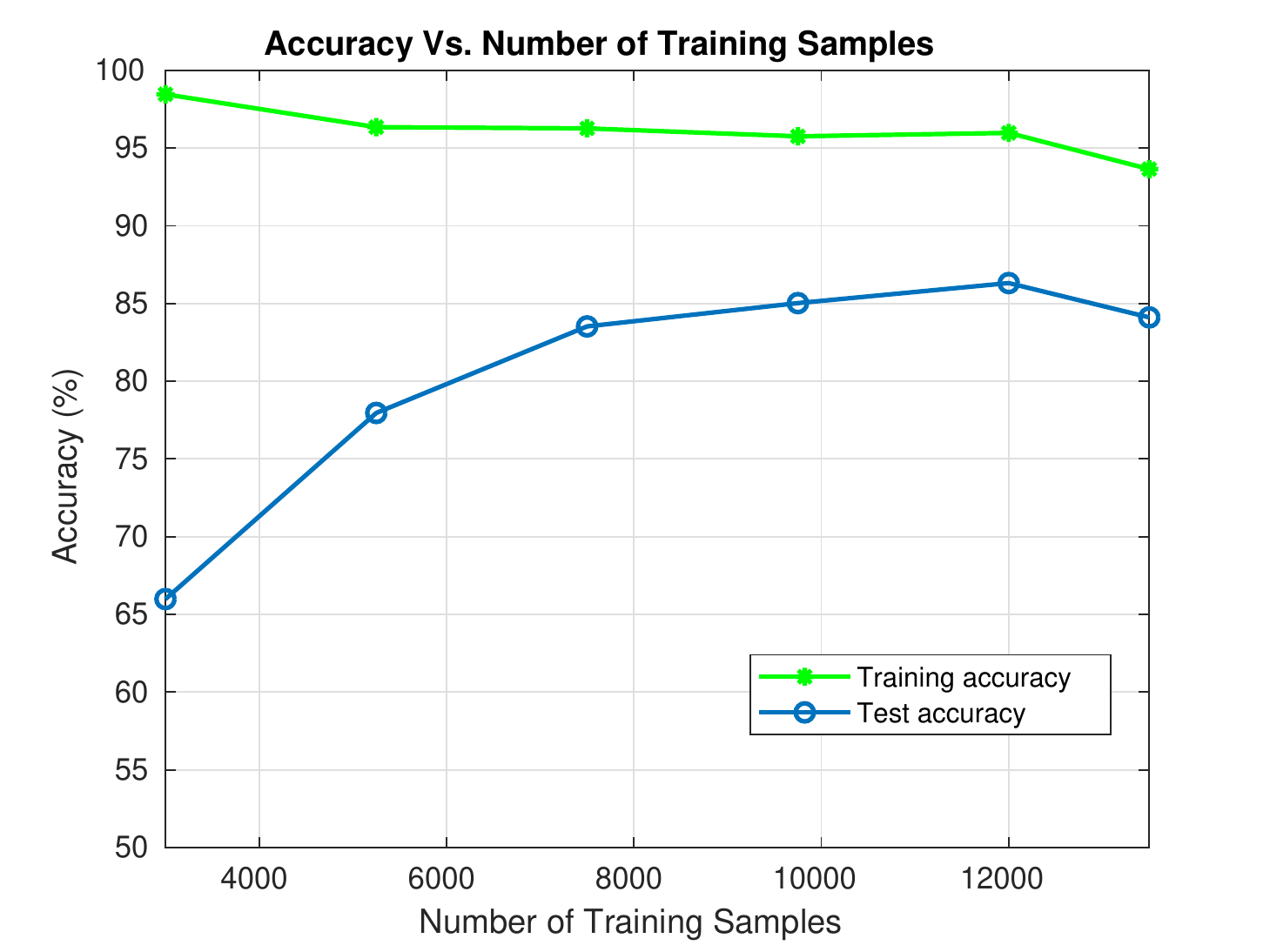}
  \caption{Test accuracy vs. number of samples.}\label{fig:5}
\endminipage\hfill
\end{figure}

After pre-training, we stack all encoders of the AE along with the Softmax layer and fine-tune network with our training dataset. Once training is completed, we test the accuracy of our model using the testing dataset. To find the optimal number of hidden layers for our model, we vary the number of hidden layers, i.e. the number of stacked auto-encoders and calculate test accuracy. Fig.~\ref{fig:4} shows the test accuracy of our DNN model vs. the number of hidden layers. From this figure, we can see that at first the accuracy increases with the number of hidden layers and then it starts to decrease. More hidden layers in a neural network means it can learn more features. As the number of hidden layer increases, the model starts also to learn the irrelevant features (noise) of the training data. Then the model is not able to perform well on the new examples and the test accuracy deteriorates. The model overfits the training data and this situation is commonly known as {\em overfitting}.

%

We achieve the maximum test accuracy of $86.31\%$ for $4$ hidden layers and slightly less test accuracy of $86.14\%$ for $3$ hidden layers. The complete DNN model along with its number of neurons per layer is shown in Fig.~\ref{fig_deep1}.  Next, we vary the number of training samples and observe its effects on both training accuracy and test accuracy.  Note that training accuracy refers to the accuracy we observe when we applying the model to the training data. 

\subsubsection*{Impact of the number of training samples on prediction accuracy} From Fig. 4, we observe that the test accuracy is low for a small number of training samples and the accuracy increases gradually with the training examples. On the other hand, the training accuracy is high for a small number, and the accuracy decreases slowly with the training examples. Initially, for a small number of training examples, the DNN learns very few distinguishing features and this enough to express the input-output relationship of the training data. This is why the training accuracy is high for a few training samples. However, for testing, the learned features are not sufficient enough to predict the output correctly. With the increase of training samples, the DNN model extracts more abstract features from the data, i.e. the model learns more about the relationship between the inputs and outputs. The testing accuracy keeps increasing with the training examples, and after some specific training examples, it starts to saturate. The reason is that as we continue to increase the training samples size there are no new distinguishing features to be learned and thus the prediction accuracy may not be increased further. But, the training accuracy keeps decreasing because the model also learns the common features (noise) of the data which ultimately degrades the performance of the model. Eventually, with a further increase in training data, the model starts to overfit the data leading to reduced test accuracy.


\section*{Conclusion}
We have presented a deep supervised learning approach to solve the  sub-band and power allocation problem in multi-cell networks.  Simulation results show that the prediction accuracy increases with the size of data samples and the number of hidden layers. However, continuous increase of the number of hidden layers  will not improve the accuracy significantly and in some cases the model may even start to learn noisy features. 
Optimal configuration of the DL model, e.g. number of hidden layers and the size of input data samples is a fundamental challenge. Also, it is crucial to develop efficient offline methods  to generate optimal/near-optimal data samples for  training the DNN architectures.  Also, since supervised deep learning is data driven, integration with cloud storage, cloud computing and SDN architecture will yield significant performance enhancement in terms of collecting and computing large number of data samples. Finally, the application of deep learning in scenarios with massive connections and spatiotemporal correlations due to  user mobility and variations in wireless channel fading characteristics would be of immediate relevance.

\bibliography{IEEEfull,references}
\bibliographystyle{IEEEtran}

\end{document}